\documentclass[aps,amsmath,amssymb,groupedaddress,twocolumn]{revtex4}
\usepackage{graphicx}

\newcommand{\nc}{\newcommand}
\nc{\beq}{\begin{equation}}
\nc{\eeq}{\end{equation}}
\nc{\beqa}{\begin{eqnarray}}
\nc{\eeqa}{\end{eqnarray}}

\def\gsim{\mathrel{\rlap{\lower4pt\hbox{\hskip1pt$\sim$}}
    \raise1pt\hbox{$>$}}}       

\begin{document}

\title{TeV gravity in four dimensions?}

\author{Xavier Calmet} \email{calmet@uoregon.edu}
\author{Stephen~D.~H.~Hsu} \email{hsu@uoregon.edu}
\affiliation{Institute of Theoretical Science \\ University of Oregon,
Eugene, OR 97403}

\begin{abstract}
We describe a model in which the fundamental scale $M_\star$  of the theory which unifies gravity and quantum mechanics  is in the TeV range, but  without requiring additional spacetime dimensions. The
weakness of gravity at low energies is due to a large vacuum expectation of a dilaton like field. The model requires a small dimensionless parameter (the self-coupling of the dilaton) but no fine-tuning. We discuss in detail the dynamical assumptions about
nonperturbative quantum gravity required within the model. We observe that $M_\star$ could be quite small, less than a TeV, and that the model could lead to copious strong coupling effects at the LHC. However, semiclassical black holes will not be produced.
\end{abstract}


\maketitle

\date{today}

Without a hierarchy, there is no hierarchy problem, which explains the appeal of models in which the fundamental scale of quantum gravity is of order TeV. The additional appeal of such models is the possibility that experiments such as LHC might directly probe the dynamics of quantum gravity by, e.g., producing microscopic black holes.

Of course, a model of TeV gravity must explain the observed weakness of gravitational effects. Models with extra dimensions \cite{ADD,RS} assume that standard model excitations are confined to a $3+1$ sub-geometry, and employ the following trick. The higher dimensional action is of the form
\begin{equation}
S=\int d^4x \, d^{d-4}x' \, \sqrt{-g} ~ \left( M_\star^{d-2} \, {\cal R} ~+~ \cdots ~~ \right) ~,
\end{equation}
so that the effective $3+1$ gravitational energy scale (Planck scale) is given by  $M_p^2 = M_*^{d-2} V_{d-4}$, where $V_{d-4}$ is the volume of the extra dimensions. By taking $V_{d-4}$ large, $M_p$ can be made of order $10^{19}$ GeV while $M_\star\sim$ TeV, at the cost of some strong dynamical assumptions about the geometry of spacetime.

In this note we describe an alternative model for TeV gravity, which does not require extra dimensions. Instead, the model contains a dilaton-like field, whose condensate determines $M_p$. The dilaton potential is assumed to allow a condensate which is much larger than the fundamental scale of the model.  Certain assumptions are required of quantum gravity in order that this possibility is realized -- in particular, concerning higher dimension operators generated by nonperturbative effects. Once these assumptions are made, and the dilaton self-coupling $\lambda$ chosen to be appropriately small, the low energy physics of the model is relatively insensitive to changes in the cutoff scale $M_*$. In this sense, the model is {\it not} fine-tuned. 

Note that it has been previously proposed to address the hierarchy problem within the framework of scalar-tensor models \cite{Cognola:2006eg}, however the idea that the scale of the fundamental theory of nature is low and potentially in the TeV region is new. One of the main differences between our approach and brane world models (in addition to the lack of extra dimensions) is that, as we shall see, in our case gravity remains weak at the TeV scale.

We assume an action of the form
\begin{eqnarray}
S&=&\int d^4x ~\sqrt{-g} ~\Big ( \frac{\phi^2}{2} {\cal R} 
~+\frac{1}{2} g^{\mu\nu}\partial_\mu\phi \partial_\nu\phi
\nonumber \\ &&  
+~ V( \phi ) ~+~ {\cal L}_{\rm sm} ~+~ \cdots  \Big ),
\label{model}
\end{eqnarray}
where ${\cal L}_{\rm sm}$ is the standard model Lagrangian and the ellipsis denote higher dimensional operators, to be discussed below. The potential for $\phi$ is chosen to be
\begin{eqnarray}
V(\phi) = - \frac{M_{\star}^2}{2} \phi^2 + \frac{\lambda}{4} \phi^4.
\end{eqnarray}
We have not introduced a $\phi^3$ coupling. This coupling does not introduce much interesting physics and could either be introduced in the potential or forbidden by imposing a parity invariance of the action. The field $\phi$ develops a vacuum expectation value $M_p$ given by 
\begin{eqnarray}
M_p=\sqrt{\frac{M_{\star}^2}{2 \lambda}}.
\end{eqnarray}
In order to reproduce Newtonian gravity, we have to identify $M_p$ with the reduced Planck scale: $M^2_p=1/(8 \pi G)=2.4353 \times 10^{18}$ GeV where $G$ is Newton's constant. The model contains two fundamental scales: the vacuum expectation value of the Higgs field and that  of the field $\phi$. The vacuum expectation value of the Higgs field is known to be 246 GeV. If the mass of the field $\phi$ is in the TeV region, there is no hierarchy problem. We shall thus assume that the fundamental scale Nature is at $M_\star \sim$ TeV. In other words we decouple the Planck scale ($\sim 10^{19}$ GeV), which from our perspective is a derived scale fixed by the vacuum expectation value of the dilaton-like field, from the fundamental scale of Nature. Whether this can be realized depends on the details of a given theory of quantum gravity (e.g., string theory \cite{Witten:1996mz,Lykken:1996fj}). In particular our model can be seen as a low energy realization of the little string theory model proposed in \cite{Antoniadis:2001sw}. In order to reproduce the Planck scale, we have to pick $\lambda \sim 8.43 \times 10^{-32}$. We regard the action (\ref{model}) as an effective theory valid up to a cutoff scale of a few $M_\star \sim$ ${\cal O}$(TeV). We shall discuss the naturalness of this choice in the sequel. Note that the idea that the Planck scale might be fixed by the vacuum expectation value of a scalar has been discussed previously in the literature \cite{Minkowski:1977aj,Zee:1978wi}. However in the scenario previously envisaged, the mass of the scalar field was close to the Planck scale and of the order of $10^{19}$ GeV.

It is worth emphasizing that, from the standpoint of naturalness, a large fundamental scale of nature and small scalar mass is very different from a small (TeV) scale of nature and large (derived) scale for gravity. In the former case, radiative corrections tend to drive the scalar mass higher -- it is unnatural for the scalar to be light. In the latter case the difficulty is in maintaining the large dilaton expectation value. Once the self-coupling $\lambda$ has been chosen small, it tends to remain small, at least in perturbation theory. To the extent that ad hoc dynamical assumptions are required (see below), they will involve nonperturbative effects and higher dimension operators. This can be compared to the assumption of a particular ad hoc (brane plus bulk) spacetime geometry in extra dimensional models.

As in  the little string theory scenario, gravity is weak in our model although the scale of the fundamental theory of gravity is assumed to be low. In little string theory, the four-dimensional Planck scale appears to be of the order of $10^{19}$ GeV although the string scale is taken to be of the order of 1 TeV. In that limit of string theory the four dimensional Planck scale is given by
\begin{eqnarray}
M_p^2=\frac{1}{g_s^2} M_s^8 V_6
\end{eqnarray}
 where $g_s$ is the string coupling constant, $M_s$ is the string mass scale and $V_6$ is the volume of the 6 extra-dimensions which are compactified. The little string theory limit corresponds to the limit $g_s \ll 1$ which corresponds to a large dilaton vacuum expectation value. The effective theory we are considering could thus be seen as an effective theory of a little string theory model and a string scale of the order of 1 TeV. 

Naively one may think that if one transforms our action to the Einstein frame, the Planck scale is decoupled from the vacuum expectation of the scalar field $\phi$. However it should be obvious that the scalar field redefinition involved in going from the Jordan frame to the Einstein frame has to preserve the vacuum of the theory. The constant which appears in the field redefinition 
\begin{eqnarray}
\phi = 2 C \exp \left (\sqrt{\frac{2}{5}} \frac{\tilde \phi}{2 C} \right)
\end{eqnarray}
thus has to be the vacuum expectation value of $\phi$  which is chosen to be the Planck scale. Furthermore in the Einstein frame, all the gauge couplings and masses are dependent on the vacuum expectation value of the field $\phi$. It would thus be incorrect to think that our model could be thought of,  in the Einstein frame, as the standard model with a strongly coupled scalar field  and a cutoff of 1 TeV: {\it in the Einstein frame all the couplings and masses of the standard model are related to the vacuum expectation value of $\phi$, which is determined dynamically.}

Let us now discuss the higher dimensional operators mentioned before. The cutoff of the effective theory is in our case very low and in the TeV region. We have to make sure that higher order operators will not destabilize the potential we are considering. An effective field theory analysis suggests dangerous higher-dimensional operators 
\begin{equation}
{\cal O} \sim {1
\over M_*^{n-4}} \phi^n~~~.
\label{danger}
\end{equation}
Such operators have a strong effect on the effective potential at values $\langle \phi \rangle \sim M_P$; without fine-tuning they would shift the minimum substantially.

A careful calculation of the effective potential reveals that the operators generated by the self-interaction of the scalar field are of the form 
 \begin{equation}
 \frac{1}{M_*^{n-4}}\lambda^\frac{n}{2} \phi^n
 \end{equation}
 rather than (\ref{danger}). They are  
always suppressed by the small parameter $\lambda$ and  do not destabilize the potential of the field $\phi$.

Furthermore the operators in (\ref{danger}) will not be generated by perturbative quantum gravity. The calculation of the effective potential involves an expansion of the field $\phi$ around its vacuum expectation value $\langle \phi \rangle$ which fixes the scale for the expansion of the metric around Minkowski spacetime once the propagator of the graviton has been normalized properly:
\begin{equation}
g_{\mu \nu} = \eta_{\mu \nu} + {1 \over \langle \phi \rangle} h_{\mu \nu}
~~,
\end{equation}
so that expanding the first term in (\ref{model}) yields a properly normalized kinetic term for the fluctuation $h$. Any coupling between $h$ and standard model fields carries a factor of $\langle \phi \rangle^{-1}$. In particular, we recover standard general relativity with coupling given by the usual Newton constant $G_N \sim M_P^{-2}$. There are three terms in our Lagrangian where $\phi$ couples to the graviton. The first two couplings involve the mass of $\phi$ and the self-interaction of $\phi$ and will  lead to effective operators which are suppressed either by factors of $\lambda$ or by terms of the type $m_\phi^4/M_p^4$ and are thus small. The direct coupling of $\phi$ to ${\cal R}$ gives a contribution to the effective potential \cite{Smolin:1979ca}
\begin{eqnarray}
\frac{\Lambda^4}{32 \pi^2} \ln\left( \frac{\left(1+9\left( \frac{\phi^2}{M_p^2}-1\right)\right) \left(1+\left( \frac{\phi^2}{M_p^2}-1\right)\right)}{1+ 4\left( \frac{\phi^2}{M_p^2}-1\right) } \right). 
\end{eqnarray}
Because the momentum cutoff $\Lambda$ in loops  is chosen to be ${\cal O}(M_\star)$ these operators will not destabilize the potential of $\phi$.
One sees that the operators which are generated are of the form:
\begin{equation}
{\cal O} \sim {
\frac{M_\star^4}{M_p^{2+n}}} \phi^{n+2},~~~
\label{nodanger1}
\end{equation}
($n \ge0$ and even) and
\begin{equation}
{\cal O} \sim {
M_\star^4  \ln\frac{\phi}{M_p}},~~~
\label{nodanger2}
\end{equation}
and are indeed not dangerous.

Other potentially dangerous operators are
\begin{equation}
\frac{1}{M_*^{n}} \phi^l {\cal O}_{\rm sm}
\label{danger2}
\end{equation}
where ${\cal O}_{\rm sm}$ is a standard model operator of dimension $n + 4 - l$. These lead, through loop corrections or condensation of 
${\cal O}_{\rm sm}$, to operators of the type (\ref{danger}) and potentially to flavor changing neutral currents or  to proton decay. However, again because the scale involved in the perturbative quantum gravity calculation is  $\langle \phi \rangle$  and because the fundamental momentum cutoff  is assumed to be $M_\star$, these operators will not be generated by loops.

Another potentially dangerous set of operators are those involving the field $\phi$ and the Higgs doublet $H$ of the standard model: $\alpha_1 \phi^2 H^\dagger H$ and $\alpha_2 \phi H^\dagger H$. These operators destabilize the Higgs potential when the vacuum expectation value of $\phi$ is introduced, so we set the couplings $\alpha_1$ and  $\alpha_2$ to zero. This is technically natural as $\phi$ is a gauge singlet and these operators will not be renormalized -- the couplings $\alpha_i$ can be set to zero at one energy scale, and remain zero at other scales. Note that quantum gravity will generate an operator 
\begin{equation}
\phi^2 H^\dagger H \frac{m_\phi^2 m_H^2}{M_P^4} \log \frac{\Lambda}{m_\phi}
\end{equation}
which however will not destabilize the Higgs potential.

Similarly, operators of the type
\begin{equation}
\frac{1}{M_*^{n}} \phi^l {\cal R} ~ {\cal O}_{\rm sm}
\label{danger3}
\end{equation}
have to be forbidden. If we took ${\cal O}_{\rm sm}=H^\dagger H$ and expanded the metric around the Minkowski spacetime, we would recover a direct coupling of the field $\phi$ to the Higgs doublet. Note however that, since gravity is weak, these operators will not be generated at the perturbative level by quantum gravity.

Although the operators ({\ref{danger}),  ({\ref{danger2}) and (\ref{danger3}) will not be generated by perturbative physics, they could be generated by non-perturbative effects related to the fundamental theory with a scale of $M_\star \sim 1$ TeV.  They are not forbidden by any symmetry and should naively be generated \cite{qgrav}. Our dynamical assumption is that this will not be the case. Obviously this depends on the quantum gravity theory which would replace our effective theory at a scale $M_\star$. Note that these operators are generically a problem for theories with a scalar field which takes values much bigger than the UV cutoff of the theory (e.g., chaotic inflation).

Our dynamical assumptions can be compared to those made in TeV scale gravity models with extra-dimensions. Instead of making an assumption about the number, the shape or size of dimensions (i.e., the semiclassical geometry of spacetime), we instead assume that a certain subset of operators will not be generated at the non-perturbative level. Modulo this assumption, our model offers a solution to the hierarchy problem and is technically natural. The running of the self-coupling of the dilaton-like field is mild and the parameter $\lambda$ does not depend strongly on the cutoff.

Since the scale for perturbative quantum gravity is the derived Planck scale (i.e. $M_p=10^{19}$ GeV), quantum gravity will not affect the phenomenology of our model in the TeV range. Black holes are unlikely to be produced in collisions at the LHC -- for example, the construction proposed in \cite{bhp} no longer produces closed trapped surfaces at TeV energies. Indeed the effect of gravity on two colliding particles is weak: due to our mechanism regular matter is screened from strong gravitational effects.

It is interesting to note that  some non-perturbative operators suppressed only by $M_\star$ can be present. At TeV energies the  field $\phi$ can be strongly coupled to ordinary matter via operators containing derivatives of $\phi$ and standard model fields, for example:
\begin{equation}
{1 \over M_*^n} ( \partial^2 \phi ) {\cal O}_{\rm sm}~~.
\label{phenoOP}
\end{equation}
Such operators do not contribute to the effective potential $V( \phi )$ (they are not eliminated by dynamical assumptions made in the previous section), but do contribute to scattering at TeV energies. If the UV completion of our model is little string theory, it is a non local theory; operators of the type (\ref{phenoOP}) can be regarded as an attempt to describe some of the non-local features of the UV completion of our model.

Another class of operators which might be generated non-perturbatively involve corrections to the Hilbert-Einstein action which are suppressed by $M_\star$ only 
\begin{eqnarray}
S_{grav.\ corr.}&=&\int d^4x \sqrt{-g} \left ( M_\star^2{\cal R} + \alpha_1 {\cal R}^2 +
 \right. \\ \nonumber && \left .
+ \alpha_2 {\cal R}_{\mu\nu}{\cal R}^{\mu\nu}+
\right. \\ \nonumber && \left .
+
 \frac{\alpha_3}{M_\star^2} {\cal R}^{\alpha\beta}_{\ \ \gamma\delta} {\cal R}^{\gamma\delta}_{\ \ \sigma\xi} {\cal R}^{\sigma\xi}_{\ \ \alpha\beta} +... \right).
\end{eqnarray}
Whether these operators will be generated or not depends on the UV completion of our model. For example,  we do not expect that these operators would be generated in a little string model. If generated, these operators would be a consequence of a modification of general relativity at $M_\star$. Unfortunately it is unlikely that these corrections can lead to interesting phenomenology. However, this shows that our model is not equivalent to simply adding a new strongly interacting scalar field to the standard model and imposing a TeV cutoff.

The phenomenology of our model is very different from that of other models with TeV gravity which involve large extra-dimensions. Ordinary matter only couples  weakly to gravitons and the bounds from LEP on graviton emission are not applicable to our model. We do not have light Kaluza-Klein excitations of the graviton and the model is thus not subject to the strong astrophysical bounds on such excitations.  

Because gravity is weak, one might think that the
fundamental length of the model is related to the Planck length. A fundamental length (or limitation on the meaning of distances below are special scale) is expected from the unification of general relativity and quantum
mechanics \cite{mead}. However, if the UV completion of our model is a little string theory the situation might be more complicated. In that
case the particles of the standard model are expected to be closed
strings with a size of the order of TeV$^{-1}$. It has been known
since the work of Amati, Ciafaloni and Veneziano \cite{Amati:1988tn}
that the smallest length which can be probed in perturbative string
theory is related to the string length $\lambda_s=\sqrt{\hbar
\alpha^\prime}$ by the relation
\begin{equation}
\Delta x \sim \frac{\hbar}{q} + q \alpha^\prime \log s,\end{equation}
where $s$ is the center of mass energy. In other words the fundamental
scale which can be probed by perturbative string theory is the string
scale \cite{footnote}.

Assuming, then, that standard model particles have finite size
$M_\star^{-1}$, one can imagine that their collisions could lead to the
formation of bound states (e.g., of little strings) which potentially  mimic the
decay of a quantum (i.e., small non-semiclassical) black hole. From
the low energy point of view, we might first see what appears to be
substructure in the familiar particles of the standard model, though
further investigation would then reveal the full spectrum of the
fundamental theory.

Although we do not expect the formation of semiclassical black holes, we do expect some strong scattering effects in the TeV region due to, e.g., finite size or operators of the type (\ref{phenoOP}) which involve the coupling of the $\phi$ particles to the standard model fields. These operators will lead to strong dynamics which could resemble compositeness as discussed in, e.g., \cite{Meade:2007sz}. 

It is known that high energy cosmic ray experiments provide a bound on  the production of black holes \cite{Feng:2001ib,Anchordoqui:2001cg,Anchordoqui:2003jr,Ringwald:2001vk,Kowalski:2002gb}. In  particular AGASA places a bound on the scale for quantum gravity as a function of extra-dimensions. Although black holes do not form in our scenario, we might apply the bound derived in \cite{Anchordoqui:2001cg} to bound the strong scattering expected in this model. The limit obtained in \cite{Anchordoqui:2001cg}  implies a bound on the cross-section  
\begin{eqnarray}
\sigma_{\nu N \to BH + X} < 0.5 \frac{1}{\mbox{TeV}^2}.
\end{eqnarray}
Assuming that our cross-section for strong dynamics is: $\sigma = M_\star^{-2}$, we get  a bound: $M_\star > 1.4 \, \mbox{TeV}$. If the fundamental scale of quantum gravity is of the order of 1.4 TeV, strong scattering at the LHC would have cross-section 
\begin{eqnarray}
 \sigma_{(pp  \to  strong \ dynamics + X)}\sim 1 \times 10^7 \mbox{fb}
\end{eqnarray}
and would thus dominate the cross-sections expected from the standard model.

In conclusion, we have described a four dimensional model in which
there is no hierarchy problem. Little string theory might be an
explicit realization of such a model, although we did not explicitly
assume this to be the case. Although gravity remains weak at the highest energy scales
in the model, we expect interesting non-perturbative dynamics due to
the dilaton like field as well as the finite ${\rm TeV}^{-1}$ size of
elementary particles.

\bigskip

\emph{Acknowledgements---} We would like to thank Michael Chanowitz, Hsin-Chia Cheng, John Gunion, Christopher Lee, Markus Luty, Yasunori Nomura and John Terning for useful discussions.
The authors are supported by the
Department of Energy under DE-FG02-96ER40969.



\bigskip


\end{document}